# Electrical and Spectroscopic Properties of SiC Detectors.

Mykola Brynza, Eduard Belas, Petr Praus, Jindrich Pipek, and Roman Grill

*Abstract*– **In this paper we compare suitability of three different semi-insulating bulk silicon carbide (SiC) materials for fabrication of radiation detectors. We prepared planar sensors with various metal contact combination and characterized detector quality by the alpha spectroscopy and I-V characteristic measurements. We observed that 4H-SiC material from the II-VI company is not suitable for radiation detector fabrication due to high concentration of vanadium doping. We also present a poor charge collection efficiency of the 4H-SiC Norstel material due to high concentration of residual impurities and we evaluated low mobility-lifetime product $\mu_e\tau^{alpha} = 3.5\times10^{-7}$ cm$^2$/V from the alpha spectroscopy. We demonstrate that the 4H-SiC material from the Cree company is the best candidate for the production of radiation detectors. We evaluated $\mu_e\tau^{alpha} = 5.4\times10^{-6}$ cm$^2$/V using AuTi/SiC/TiAu contact structure.**

## I. Introduction

SEMI-insulating (SI) Silicon Carbide (SiC) is a promising material for preparation of the room-temperature radiation detectors [1-3] operating at extremely harsh environmental conditions such as in space and the automotive, aeronautic and nuclear industries. The main advantages of SiC are low leakage current due to wide band gap, high critical breakdown field that allows application of high electric fields and high radiation hardness due to the high displacement energy of atoms. One of the main problems of preparation of high performance bulk SiC detectors is still crystallographic quality of the detector-grade bulk material.

In this paper, we compare transport properties of SiC detectors prepared from the single crystalline SI wafers produced by three different vendors (Cree Inc., Norstel AB and II-VI Inc.). For detector characterization the alpha particle spectroscopy and I-V characteristics measurement were used. We evaluated Cree material as the best choice for preparation of bulk SiC radiation detectors.

## II. Experimental

Bulk SI SiC samples were prepared from wafers produced by the II-VI, Norstel and Cree companies. Planar samples with dimensions 5 mm×5 mm×0.5 mm were cut from each wafer and gold, nickel, chromium or titanium electrical contacts were prepared by thermal evaporation on both sides of the samples. Charge collection efficiency of all prepared detectors was tested by alpha spectroscopy using $^{241}$Am alpha particle source and the I-V characteristics measurement was used for characterization of contact properties. The scheme of the high impedance measurement of I-V characteristics is presented in Fig. 1.

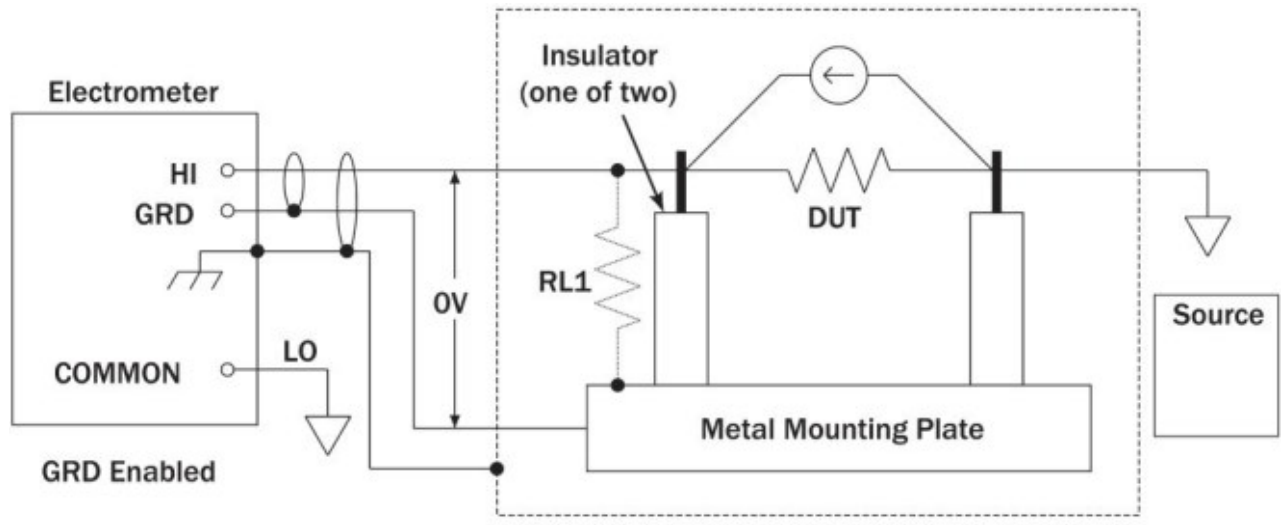

Fig. 1. The scheme of the high impedance measurement of I-V characteristics [4].

## III. Results and discussion

Table I. shows dominant chemical elements observed in used wafers using Glow Discharge Mass Spectroscopy (GDMS).

TABLE I. GDMS ELEMENT ANALYSIS OF USED SIC MATERIALS

| Element | II-VI | Norstel | Cree |
|---|---|---|---|
| B | 150 | 260 | 120 |
| Na | 130 | 280 | 210 |
| S | 180 | 630 | 410 |
| V | 430 | < 5 | < 5 |
| Cu | 320 | 760 | < 20 |
| P | 30 | 130 | < 30 |
| Zn | 100 | 310 | < 80 |

Units in the table – ppb atomic

It was found that the II-VI material additionally to the intentionally doped vanadium contains also high concentration of other impurities. The high concentration of Cu, S, P and Zn was detected in the Norstel material, as well. The Cree material occurred to be the purest ones with only higher concentration of sulfur. Charge collection efficiency (CCE) of prepared detectors with gold electrical contacts was tested for each material using alpha spectroscopy. It was observed that

Manuscript received December 13, 2019. This work was supported by the Grant Agency of the Czech Republic under Grant No. 18-12449S .

M. Brynza, E. Belas, P. Praus, J. Pipek, and R. Grill are with the Charles University, Faculty of Mathematics and Physics, Institute of Physics, Ke Karlovu 5, Prague 2, CZ-12116, Czech Republic (e-mail: brynza.mykola@karlov.mff.cuni.cz, eduard.belas@mff.cuni.cz, praus@karlov.mff.cuni.cz, jindrich.pipek@gmail.com, grill@karlov.mff.cuni.cz).

the II-VI detectors are not able to detect alpha particles most probably due to vanadium doping because it is known that vanadium creates deep recombination centers for electrons in SiC and suppress charge collection efficiency [5-6]. In Fig. 2 we show the pulse height alpha spectrum of the Norstel detector and the bias dependence of CCE fitted by the single carrier Hecht equation [7] presented in the inset. We evaluated the mobility-lifetime product $\mu_e\tau_e^{alpha} = 3.5\times10^{-7}$ cm$^2$/V. Linear bias dependence of CCE shows very low detection efficiency, which is most probably caused by the high concentration of residual impurities in the material.

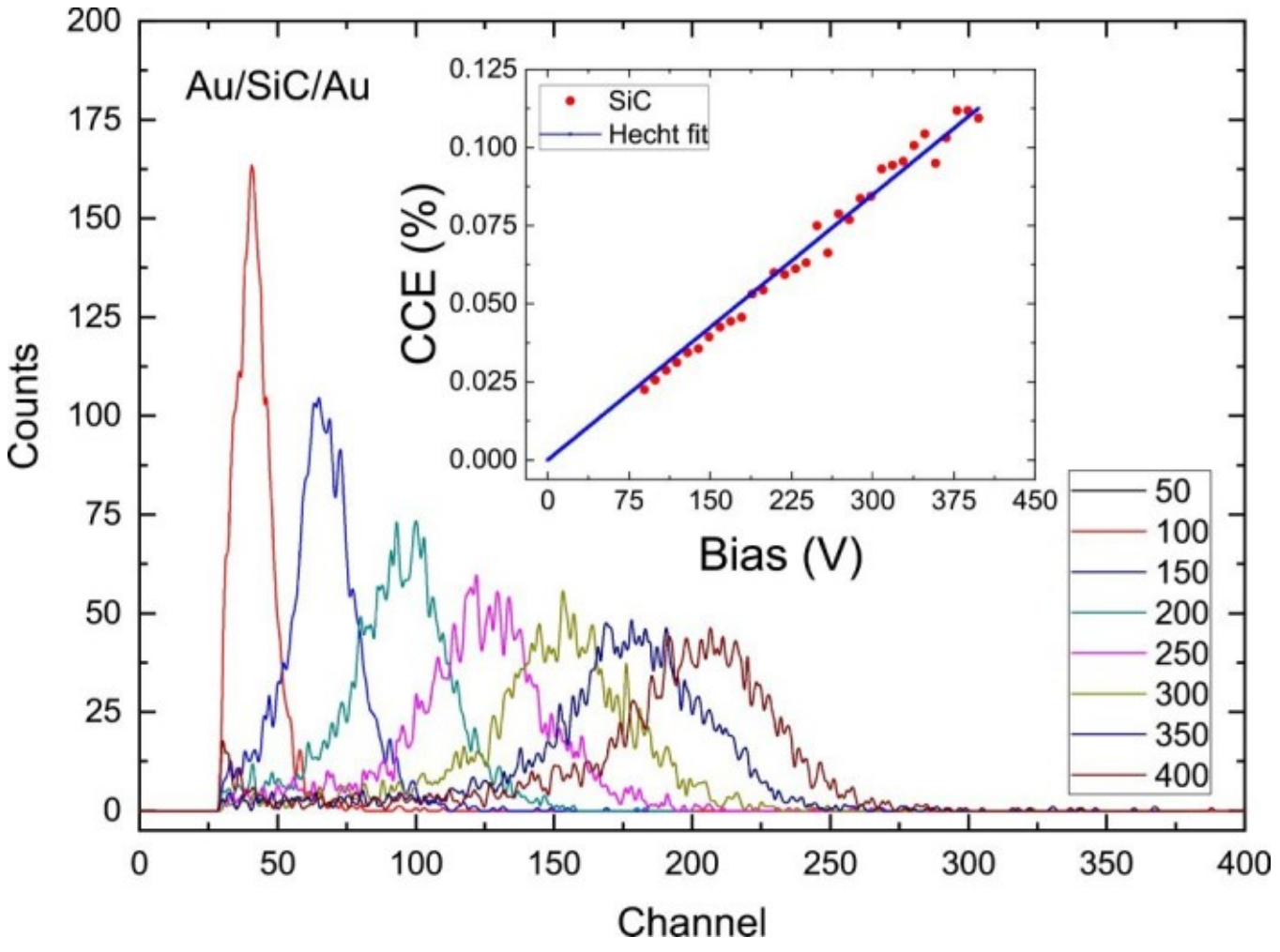

Fig. 2. Alpha particle pulse height spectrum of the Norstel detector with Au/SiC/Au contact structure. The bias dependence of CCE and Hecht fit are presented in the inset.

Pulse height alpha spectrum of the Cree detector with gold contacts is shown in Fig.3 and the bias dependence of CCE fitted by the single carrier Hecht equation is presented in the inset. We evaluated the mobility-lifetime product $\mu_e\tau_e^{alpha} = 4.1\times10^{-6}$ cm$^2$/V. Generally, we observed that the Cree detectors shows much higher CCE than other detectors. Therefore we concentrated our further investigation of the electrical contact properties to the Cree material only.

Due to poor adhesion of evaporated thin gold contacts, we managed to prepare various combinations of metal contacts. We observed that the combination of Au contact with thin layers of Ni, Cr or Ti highly increases contact adhesion with no significant influence to the ability of alpha particles detection. Comparison of I-V characteristics of the Cree detector with Au/SiC/Au and Ni/SiC/Ni contact structures is presented in Fig.4. We demonstrate that Au/SiC/Au contact structure significantly decreases leakage currents and we calculated series resistance $R_S$(Au)= $9.1\times10^{12}$ Ω and $R_S$(Ni)= $3.7\times10^{14}$ Ω for Au and Ni contacts, respectively.

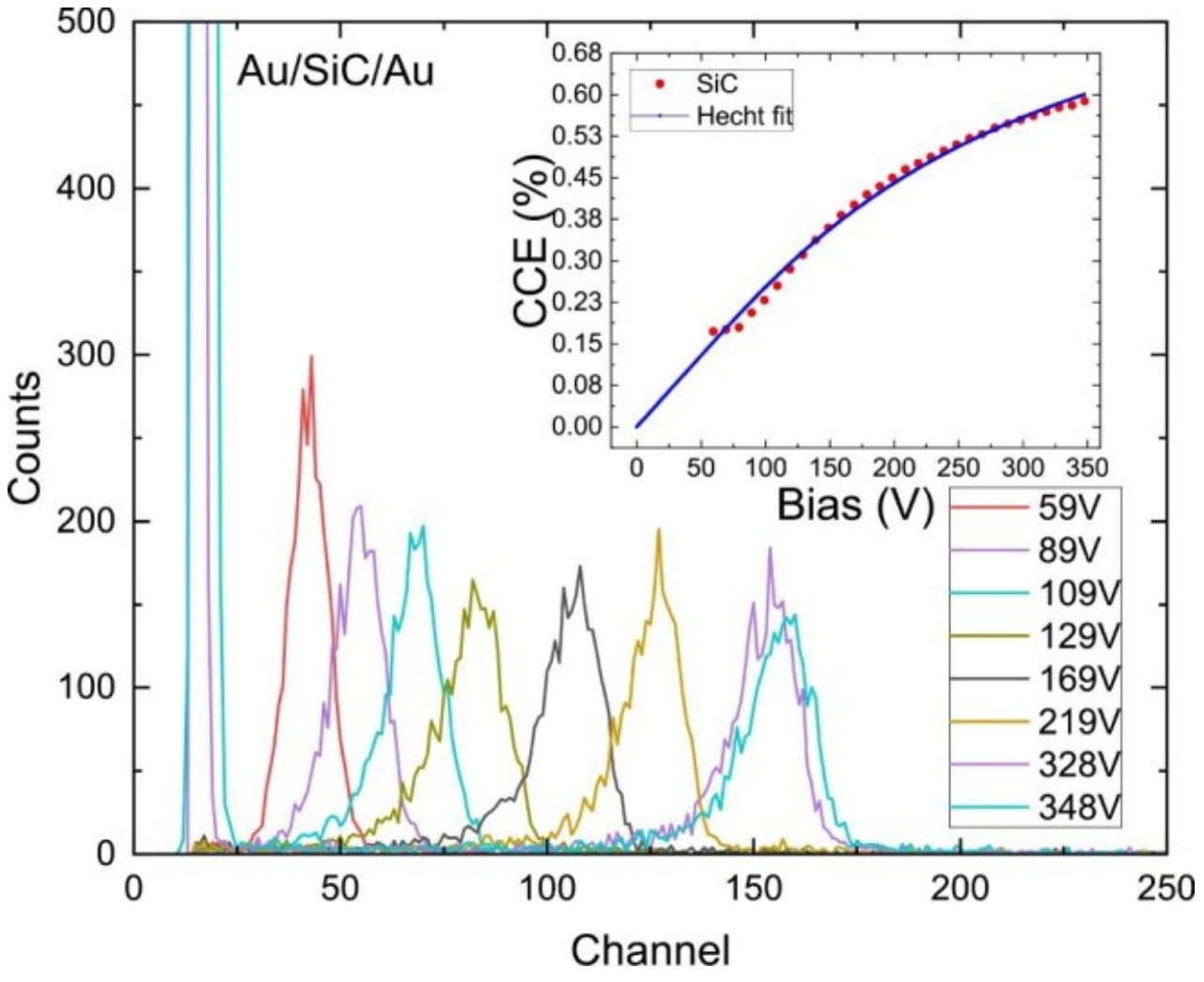

Fig. 3. Alpha particle pulse height spectrum of the Cree detector with Au/SiC/Au contact structure. The bias dependence of CCE and Hecht fit are presented in the inset.

We also found that the adhesion of even thin simple Ni contacts is much higher than of Au contacts. Nevertheless, increasing thickness of the contact reduce noises in the high bias measurements.

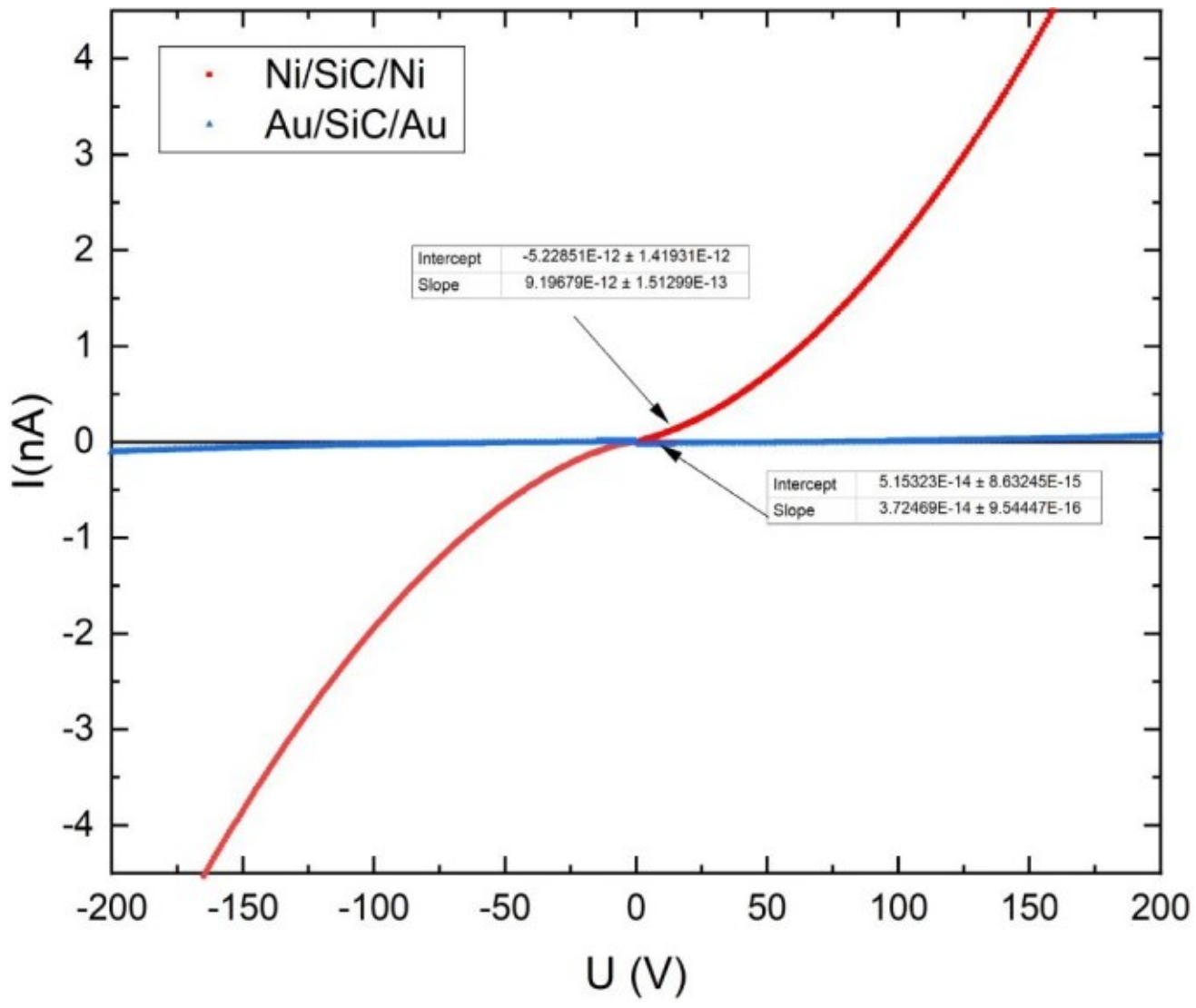

Fig. 4. I-V characteristics of the Cree detector with Au/SiC/Au and Ni/SiC/Ni metal contact structures.

The pulse height alpha spectrum of the Cree detector with Ni/SiC/Ni contacts is shown in Fig.5. We calculated the mobility-lifetime product $\mu_e\tau_e^{alpha} = 4.6\times10^{-6}$ cm$^2$/V from the bias dependence of CCE presented in the inset of Fig.5, which is similar value obtained for Au/SiC/Au detector. The highest value of mobility-lifetime product was evaluated for Cree detector prepared with Au-Ti/SiC/Ti-Au contact structure. Fig.6. shows the pulse height alpha spectrum of the Cree detector with Au-Ti/SiC/Ti-Au contact structure using 70 nm of Au and 5 nm of Ti, where $\mu_e\tau_e^{alpha} = 5.4\times10^{-6}$ cm$^2$/V was evaluated.

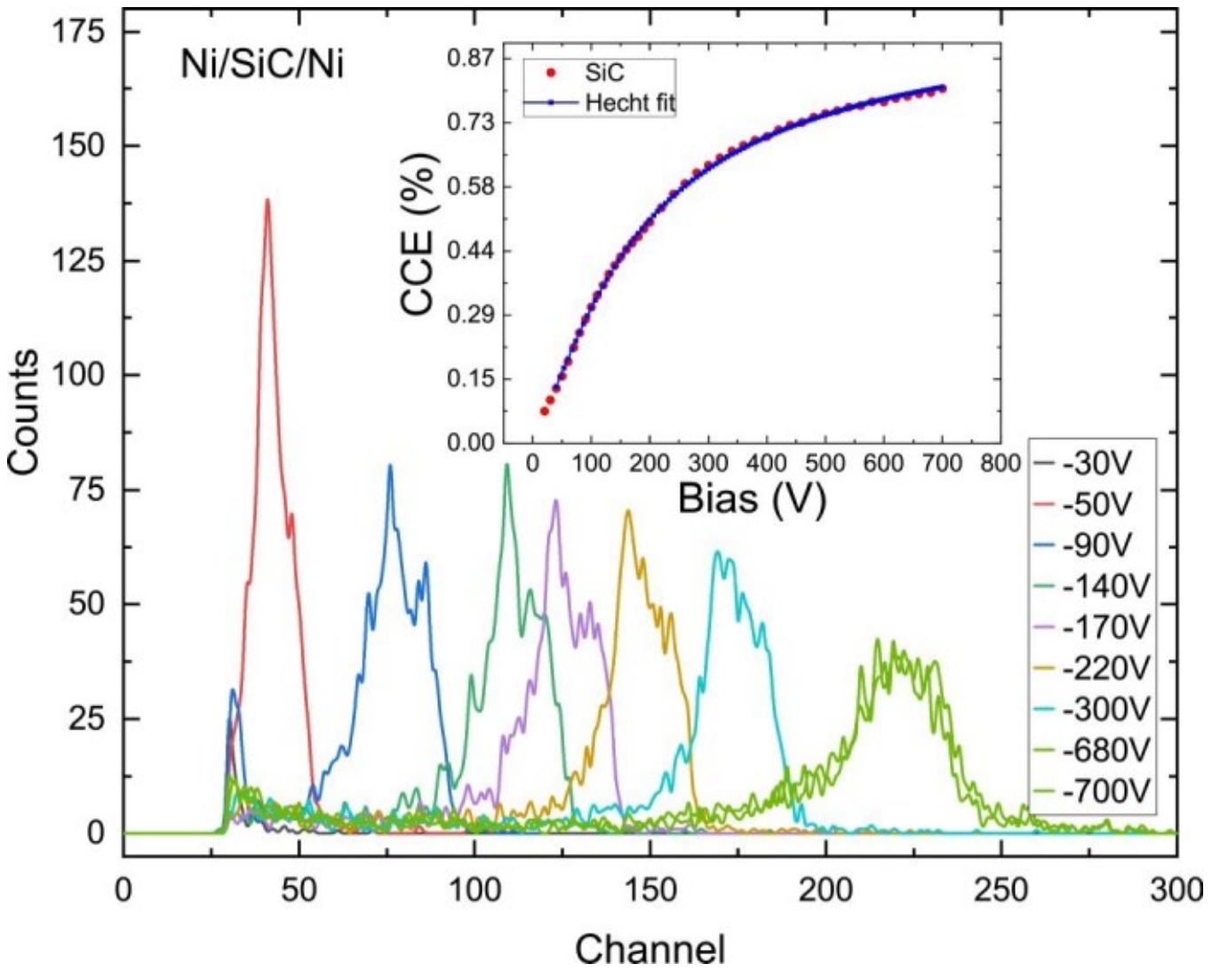

Fig. 5. Alpha particle pulse height spectrum of the Cree detector with Ni/SiC/Ni contact. The bias dependence of CCE and Hecht fit are presented in the inset.

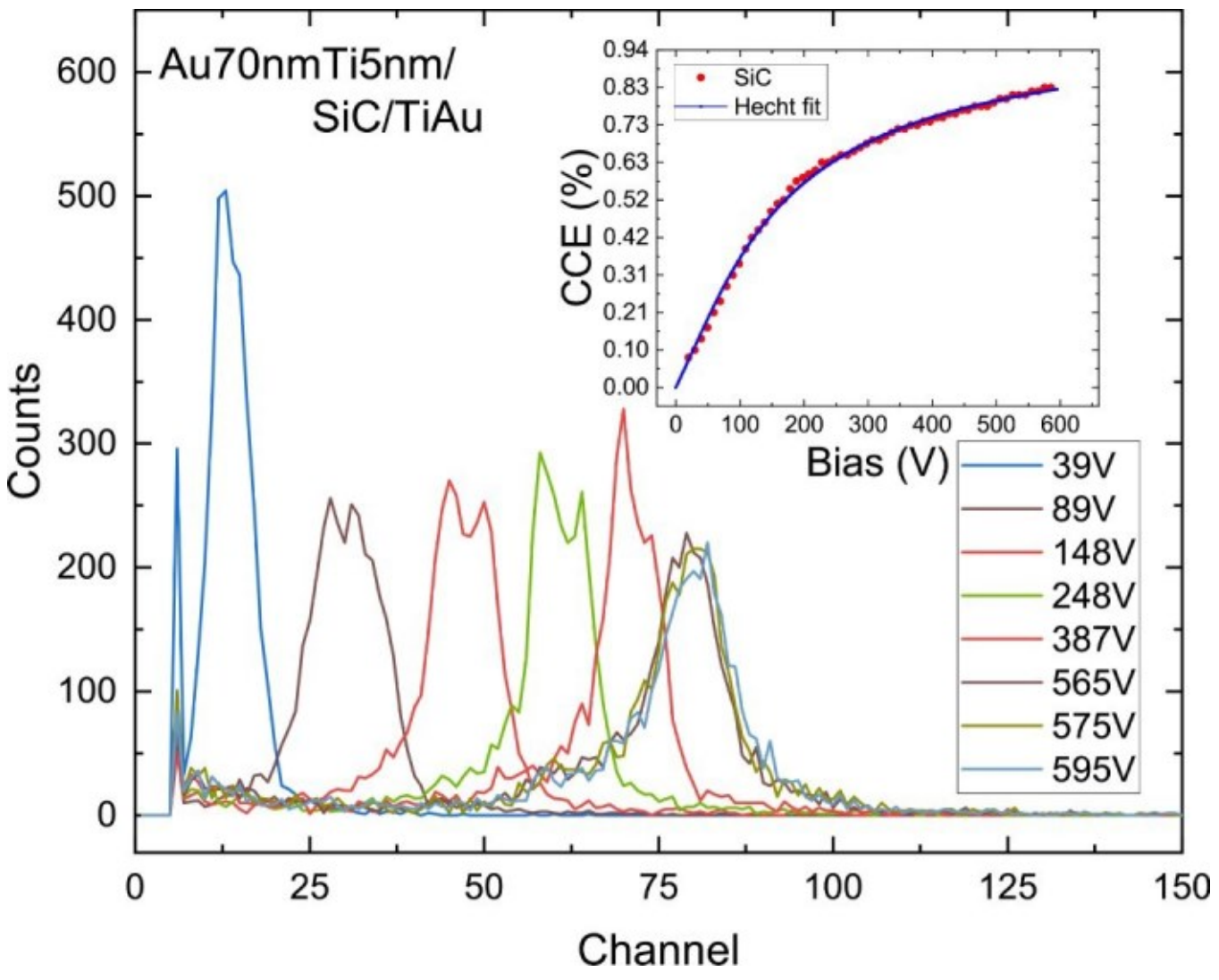

Fig. 6. Alpha particle pulse height spectrum of the Cree detector with AuTi/SiC/TiAu contact structure. The bias dependence of CCE and Hecht fit are presented in the inset.

## IV. Conclusion

Suitability of semi-insulating bulk SiC materials prepared by three different producers for the fabrication of radiation detectors has been compared. We demonstrated that the 4H-SiC material from the Cree company is the best candidate for the production of the bulk radiation detectors. We evaluated mobility lifetime product $\mu_e\tau_e^{alpha} = 5.4\times10^{-6}$ cm$^2$/V using AuTi/SiC/TiAu contact structure. In the case of the 4H-SiC Norstel material we found a poor charge collection efficiency due to high concentration of residual impurities. We also observed that SI 4H-SiC material produced by the II-VI company is not suitable for radiation detector fabrication due to high concentration of vanadium doping.